\documentclass[
superscriptaddress,
twocolumn,
showpacs,
bibnotes,
amsmath,
amssymb,
aps,
prx,
floatfix,
]{revtex4-2}

\usepackage{blindtext}
\usepackage{graphicx}
\usepackage{svg}
\usepackage{gensymb}
\usepackage{lipsum}  
\usepackage{multirow}
\usepackage{xcolor}
\usepackage{enumitem}

\begin{document}
\title{Magnetocrystalline Anisotropy and 3D Hopping Conduction at the Surface of FeSb$_2$}

\author{Jarryd A. Horn}
    \affiliation{Maryland Quantum Materials Center and Department of Physics, University of Maryland, College Park, Maryland 20742, USA}

\author{Yun Suk Eo}
    \affiliation{Maryland Quantum Materials Center and Department of Physics, University of Maryland, College Park, Maryland 20742, USA}
    \affiliation{Texas Tech University Department of Physics, Lubbock, Texas 79409, USA}

\author{Keenan Avers}
    \affiliation{Maryland Quantum Materials Center and Department of Physics, University of Maryland, College Park, Maryland 20742, USA}
    
\author{Hyeok Yoon}
    \affiliation{Maryland Quantum Materials Center and Department of Physics, University of Maryland, College Park, Maryland 20742, USA}

\author{Ryan G. Dorman}
    \affiliation{Maryland Quantum Materials Center and Department of Physics, University of Maryland, College Park, Maryland 20742, USA}
    \affiliation{University of Colorado Department of Physics, Boulder, Colorado 80309, USA}

\author{Shanta R. Saha}
    \affiliation{Maryland Quantum Materials Center and Department of Physics, University of Maryland, College Park, Maryland 20742, USA}
    
\author{Johnpierre Paglione}
    \affiliation{Maryland Quantum Materials Center and Department of Physics, University of Maryland, College Park, Maryland 20742, USA}
    \affiliation{Canadian Institute for Advanced Research, Toronto, Ontario M5G 1Z8, Canada}
    
\date{\today}
    
\begin{abstract}
 Motivated by the recent discovery of metallic surface states in the $d$-electron Kondo insulator candidates FeSi and FeSb$_2$, along with some recent reports of magnetic correlations in the surface transport properties of FeSi, we have investigated the low temperature surface magnetotransport properties of FeSb$_2$. By using a Corbino disk transport geometry, we were able to isolate the electrical transport properties of a single surface of our samples and study the [110] and [101] naturally forming faces separately. Studying the relationship between the applied magnetic field, current direction and crystal symmetry has allowed us to separate possible contributions to the magnetotransport anisotropy. Unlike previous studies of SmB$_6$ surface states, we find no two-dimensional Drude-like dependence on field orientation relative to current direction, but instead a magnetocrystalline anisotropy that appears to originate from local moment scattering with a well defined easy-axis along the [100] direction. We compare these results with the magnetotransport properties of the conducting surface states on the [111] facet of FeSi.  We also find evidence of 3D variable-range hopping conduction below the bulk-to-surface crossover, extending below 1 K, which implies that the electrical transport at the surface of these materials is carried by a thin, but 3D conducting channel, which is inconsistent with the lower dimensional states expected for a strong topological insulator. 
\end{abstract} 

\maketitle

\section{Introduction}

Until recently, the study of Kondo insulator (KI) physics have been focused on materials with heavy $f$-electron orbitals near the Fermi level \cite{Dzero2016}. In typical KI systems, the local $f$-moment hybridizes with conduction electrons to form a narrow gap, with well studied examples being SmB$_6$ and YbB$_{12}$ \cite{Menth1969,Martin1979,Rosler2014,Takabatake1998,Weng2014,Kang2016}, and more recently in U$_3$Bi$_4$Ni$_3$ \cite{Broyles2025}. In these particular systems, a clear plateau in the low temperature electrical resistance is found to arise due to residual conduction from surface states that electrically short conduction from bulk activated charge carriers \cite{Eo2019}. While angle-resolved photoemission spectroscopy (ARPES) and scanning tunneling microscopy quasi-particle interference (STM-QPI) studies on clean, in-situ cleaved surfaces show evidence of these gapless surface states, the topological classification of these surface states are still heavily debated \cite{Matt2020,Hlawenka2018}.

First reports of the Kondo insulator descriptions of FeSi and FeSb$_2$ came from examining the bulk magnetization of these materials and the apparent activation of a singlet ground state to a high temperature paramagnetic state, consistent with singlet formation in Kondo insulators \cite{Petrovic2005,Schlesinger1997}. Speculation of topological Kondo insulator physics arose after the first reports of surface conduction in FeSi by dimensional scaling of resistance \cite{Fang2018}, followed by spectroscopic evidence of gapless surface states reported in FeSb$_2$ and FeSi by ARPES \cite{Xu2020,Changdar2020} and later demonstrated unambiguously by non-local transport \cite{Eo2023}. Subsequent magnetoresistance measurements on FeSi showed that the surface states are magnetically correlated with evidence of magnetically ordered surface states \cite{Ohtsuka2021,Breindel2023,Deng2023,Avers2024}. Recent measurements of surface magnetotransport anisotropy in FeSb$_2$ suggest similarities between SmB$_6$ and FeSb$_2$ but do not report similar correlated physics as was found in FeSi \cite{Eaton2024}, although theoretical calculations suggests that FeSb$_2$ is an incipient altermagnet which may stabilize order with doping \cite{Mazin2021}.

The topological KI description of FeSb$_2$ is contested by spectroscopic and ab-initio studies \cite{Chikina2020,Li2024}. For example, Li {\it et al.} note that spectroscopic evidence support the presence of surface states on the [010] and [110] surfaces with the absence of surface states on [001] surfaces of FeSb$_2$ \cite{Li2024}. The authors also propose an alternative theory to the Kondo insulator description of FeSb$_2$ involving an excitation of the bulk spin state across a multiplet gap. Another possible mechanism put forth by Chikina {et al.} and supported by ab initio calculations to match ARPES data is the possible existence of structural distortions at the surface, which would depend on how the surface is terminated \cite{Chikina2020}.

In this paper, we report on low temperature transport properties of the [110] and [101] surfaces of FeSb$_2$, demonstrating the crossover in magnetotransport anisotropy from the bulk-dominated conduction regime at high temperatures to the surface-dominated conduction regime at low temperatures by comparing the angle dependence of the MR. A detailed analysis of the magnetic field dependence of the low-temperature surface transport is shown to be consistent with local moment scattering, and comparisons to similar measurements on the [111] surface of FeSi indicate that magnetic correlations observed in the ferromagnetic surface states of FeSi may play a similar role on the surface of FeSb$_2$.

\begin{figure}[h!t!]
    \centering
    \includegraphics[width=0.95\linewidth]{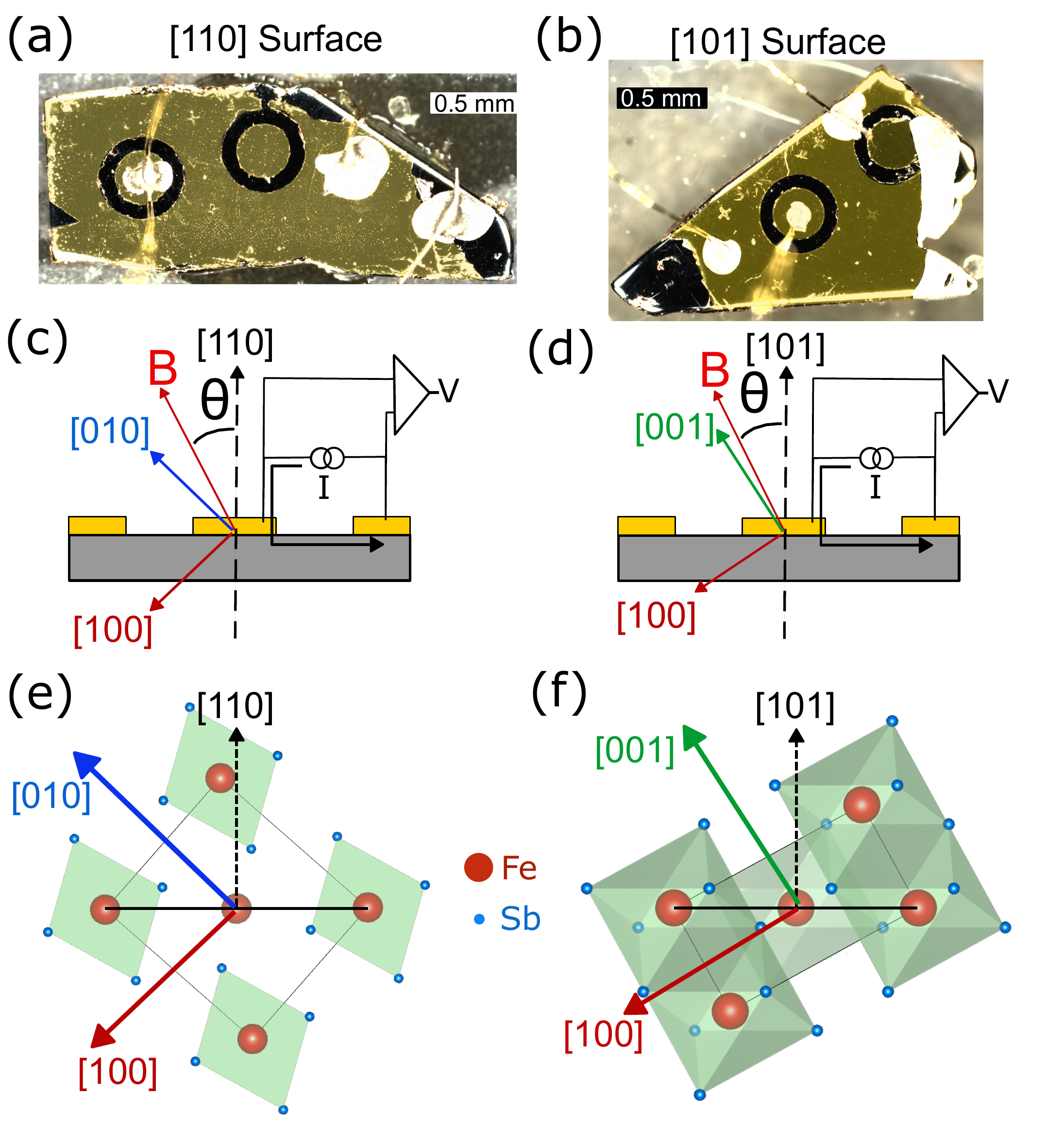}
    \caption{(a-b) FeSb$_2$ transport samples patterned with Corbino disk transport geometries deposited with Ti/Au on [110] surface and  [101] surface. (c-d) Schematic of magnetic field rotation in- and out-of-plane of [110] surface a-b plane and [101] surface a-c plane. (e-f) Schematic of a-b plane and a-c plane aligned with the orientation of (c-d).}
    \label{fig:overview}
\end{figure}

\section{Experimental Details}

While all other previously reported topological Kondo insulator candidates, including FeSi, have cubic crystal symmetry, FeSb$_2$ crystallizes with orthorhombic crystal symmetry \cite{structure_FeSb2} of space group $Pnnm$ with unit cell dimensions of $a$=5.834 \AA, $b$=6.530 \AA, and $c$=3.193 \AA. This complicates the study of surface states in this material, which entails a larger permutation space for surfaces and unique field directions. For this study, we will focus on the study of [110] and [101] surfaces because of the convenience of being provided with clean facets of these surfaces from crystal formation.

In order to carefully study the magnetotransport  properties of the surface of FeSb$_2$, we obtained single-crystal samples with as-grown [110] and [101] facets. We pattern Corbino disks on those facets by standard photolithography followed by e-beam evaporation of Ti/Au (30A/1500A), as shown in Fig.~\ref{fig:overview} (a) and (b). These metalized patterns not only confine the current path, but also provide excellent electrical contacts. This technique probes only a single surface and therefore provides a way to rotate the magnetic field in and out of the plane in which our current is confined \cite{Eo2018}. When applied to a 2D electron gas (2DEG), this method allows an accurate measurement of the 2D transport parameters, as has been shown with the surface states of SmB$_6$ \cite{Eo2020}. The FeSi sample used in this paper is the same four-terminal [111] surface Corbino sample reported in previous reports \cite{Eo2023} and \cite{Avers2024}.

Magnetoresistance (MR $\equiv 100\%*(R(B)-R(0))/R(0)$, for resistance, $R$, at magnetic field, $B$) was measured using Quantum Design Dynacool with built in resistance bridge using 1 $
\mu$A excitation. The direction of the field was controlled using a horizontal (single axis) rotator insert with in-plane and out-of-plane rotator pucks. Resistance data below 1.8 K was measured in a Quantum Design PPMS using an adiabatic demagnetization accessory and built-in resistance bridge using 100 nA excitation.

\section{Surface Corbino Disk Angular Magnetoresistance}

\begin{figure}[h!]
    \centering
    \includegraphics[width=0.95\linewidth]{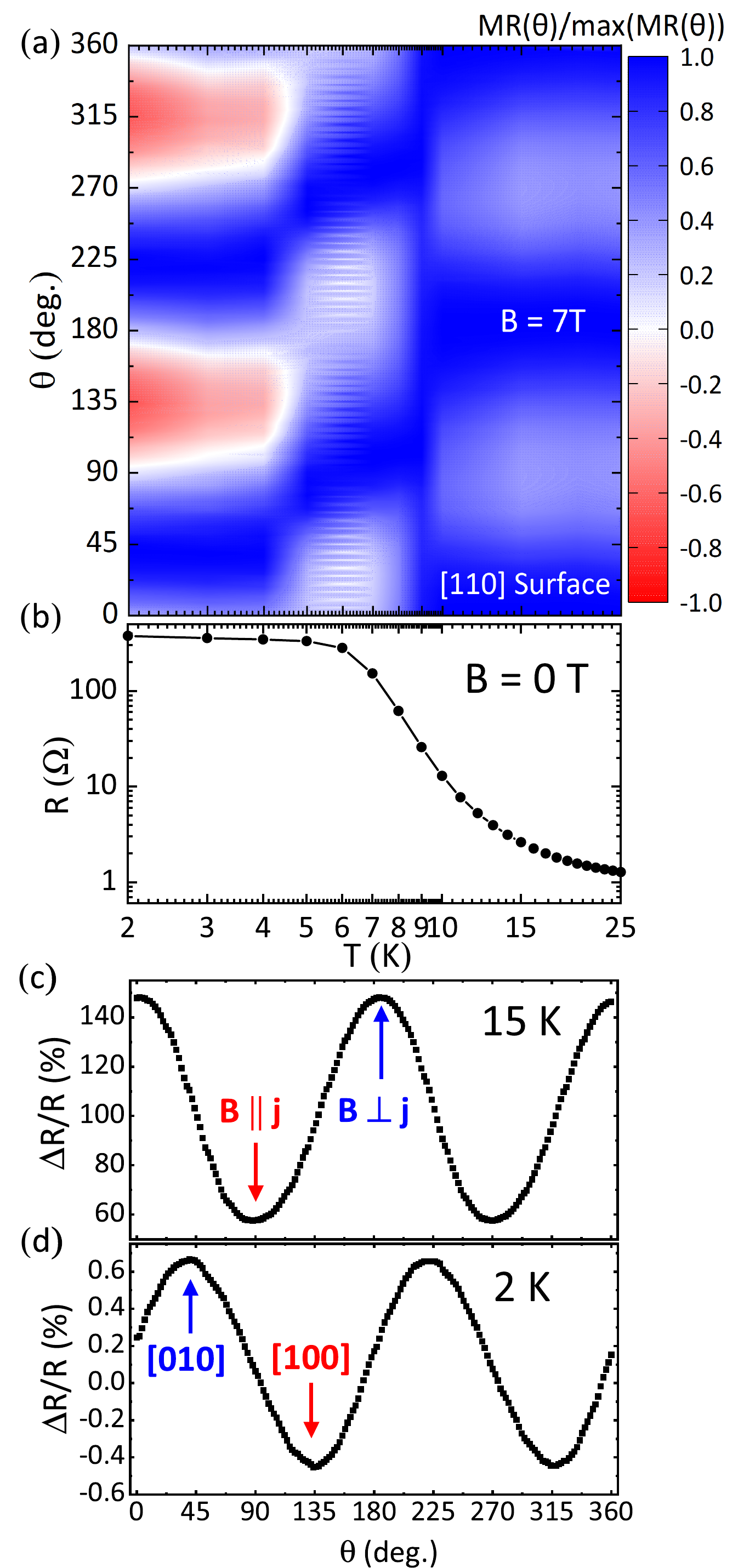}
    \caption{Crossover in angular magnetoresistance symmetry as shown in the normalized angle dependence of MR from 25K down to 2K for current in [110] surface with field swept in a-b plane. (a) Angle sweep vs temperature color-plot of magnetoresistance normalized to the maximum magnetoresistance for each isotherm angle sweep at 7 T magnetic field with angle sweep depicted in Fig\ref{fig:overview}c. (b) Zero field resistance showing plateau on cooling coinciding with symmetry change in angular magnetoresistance and crossover from bulk to surface dominated transport regime according to inverted resistance measurements \cite{Eo2023}. (c-d) 7 T magnetoresistance angle sweep at 15 K and 2 K corresponding to vertical line cuts of color-plot in (a) without normalization.}
    \label{fig:AMR}
\end{figure}

\begin{figure*}
    \centering
    \includegraphics[width=\textwidth]{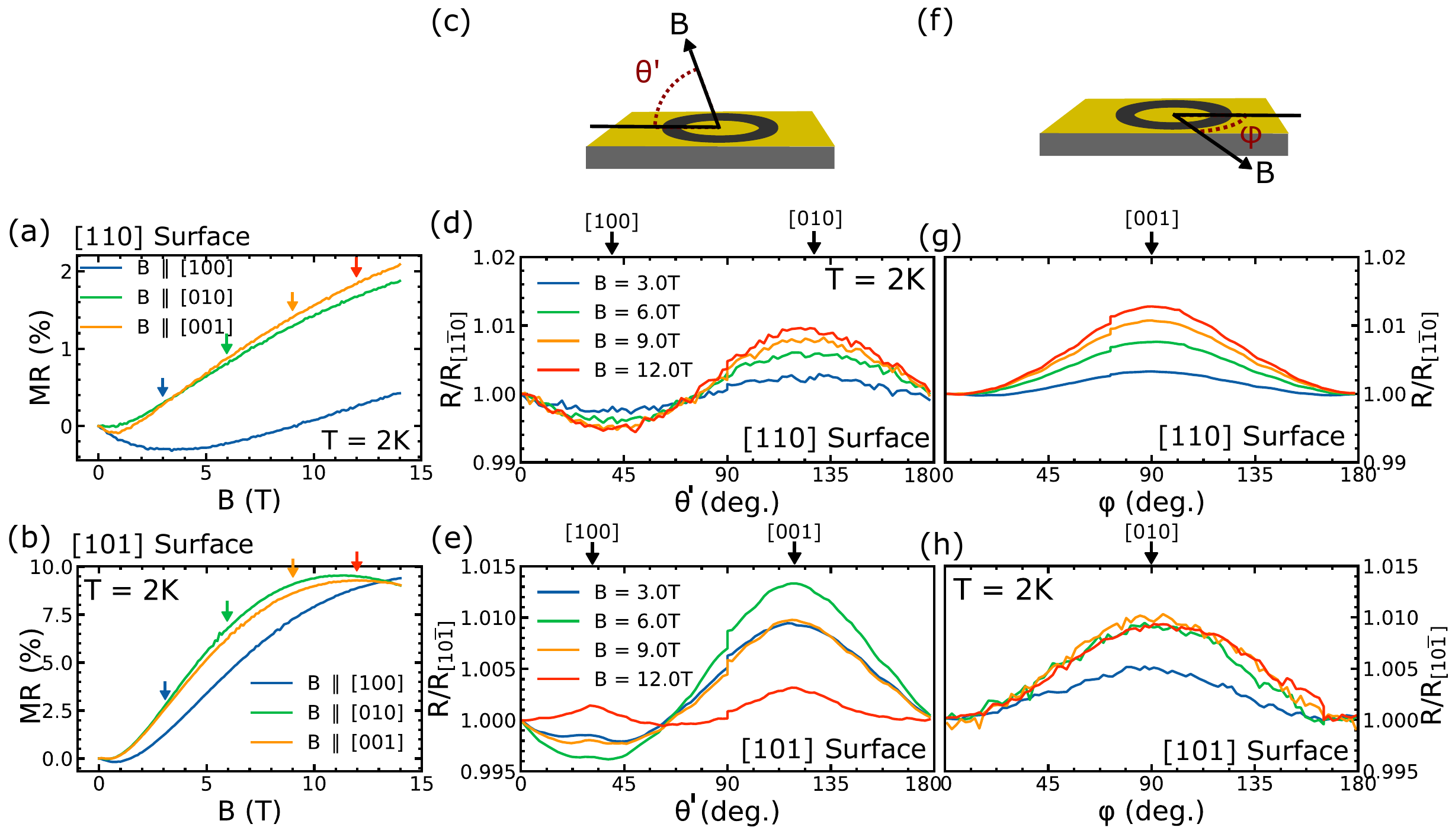}
    \caption{(a-b) Magnetic field dependence for field applied along the principle crystal axes at 2K for both current directions, [110] and [101], with arrows corresponding to fields at which angle sweeps were measured for fixed magnetic field. (c-e) Fixed field angle sweeps out-of-plane of the surface with resistance normalized to the resistance for field applied along the corresponding in-plane magnetic field direction and angle shifted by 90$\degree$ such that $\theta=180\degree$ $\phi=0\degree$ correspond to the same direction and normalized value of unity. (f-h) Fixed field angle sweeps in-plane of the surface.}
    \label{fig:B-sweeps}
\end{figure*}

Measurements of the angle dependence of MR show three distinct regions of angular response that map well to temperature dependent changes in the zero-field resistance, as shown in Fig.~\ref{fig:AMR}. The correlation between the crossover in the angle dependence in MR with the plateau in low temperature electrical resistance associated with surface states confirms that these regions are given by a low temperature surface MR ($T\leq4 K$), high temperature bulk MR ($T\geq10 K$) and intermediate mixed behavior. 

To visualize the change in the angle dependence, we normalized the MR by the maximum value of MR$(\theta)$ for each angle sweep at constant temperature and field. At high temperatures, above the resistance plateau, the electrical transport is predominantly through the bulk. In this regime, when magnetic field is applied in-plane of the Corbino disk, the component of MR attributable to Lorentz force is at a minimum since the component of the current perpendicular to the magnetic field is at a minimum. This is reflected in Fig.~\ref{fig:AMR} (a) and (c) for temperatures above 8 K by roughly 50\% reduction in MR for field along current direction in comparison to field fully perpendicular to current. At temperatures below 5 K, in contrast, results from surface angular MR reveals a strong dependence on field direction relative to the principle crystallographic axes, rather than the 2D current direction, as depicted in Fig.~\ref{fig:AMR} (d). In this temperature range, the in-plane and out-of-plane MR are comparable and therefore inconsistent with suppression of Lorentz force on charge carriers. This result seems to rule out the Drude contribution expected from a 2DEG from playing a significant role in the surface transport properties of FeSb$_2$. 

The low-temperature MR field dependence of two corbino samples with [110] and [101] surface geometries is presented in Figs.~\ref{fig:B-sweeps}a and b, respectively, for field orientations pointed along each principal crystal axis. The 2~K field response is characterized by a negative MR at low fields for all cases, most pronounced for fields along [100], and a positive MR for larger  fields. Near 10~T, the [101] surface sample MR enters a high field region characterized by a growing negative MR contribution which may be the reprisal of the low field negative MR due to the saturation of the positive MR component.

To understand the angle dependence of these samples, we compare magnetic field rotations out of the plane of the surface (Fig.~\ref{fig:B-sweeps}d and e) and magnetic field rotations in the plane of the surface (Fig.~\ref{fig:B-sweeps}g and h). In order to align visually the angles for in-plane and out-of-plane angle sweeps at 180$^o$, we define $\theta'=90-\theta$ and normalize to the common direction between in-plane and out-of-plane angle sweeps. For example, $R/R_{[1\bar10]}(\theta=180^o)=R/R_{[1\bar10]}(\phi=180^o)=1$ for the [110] surface and $R/R_{[10\bar1]}(\theta=180^o)=R/R_{[10\bar1]}(\phi=180^o)=1$ for the [101] surface. The [100],[010] and [001] directions are annotated on the plots for Fig.~\ref{fig:B-sweeps}d-h for clarity. For both surfaces and all field values, the amplitude of the angle dependence for field sweeps out-of-plane (Fig.~\ref{fig:B-sweeps}d-e) and in-plane (Fig.~\ref{fig:B-sweeps}g-h) are comparable in magnitude with local maxima for field along [010] and [001] field directions. This further highlights the lack of dependence of the MR on current direction and rather that there is no shift in symmetry away from the crystalline anisotropy spanning from low fields (where the negative MR component is most prominent) to high fields (where the positive MR component is most prominent). For example, for a 2D system in which there is a negative MR contribution due to spin-orbit coupled magnetic fluctuations and as positive MR contribution due to Lorentz force on free carriers, we would expect that the MR anisotropy would shift from crystalline anisotropy at low fields to a high field dependence on the field relative to current direction. Such a scenario is not consistent with our observations. 

\section{Comparison of [010] vs [001] Surface with uniaxial transport}

\begin{figure*}
    \centering
    \includegraphics[width=0.9\linewidth]{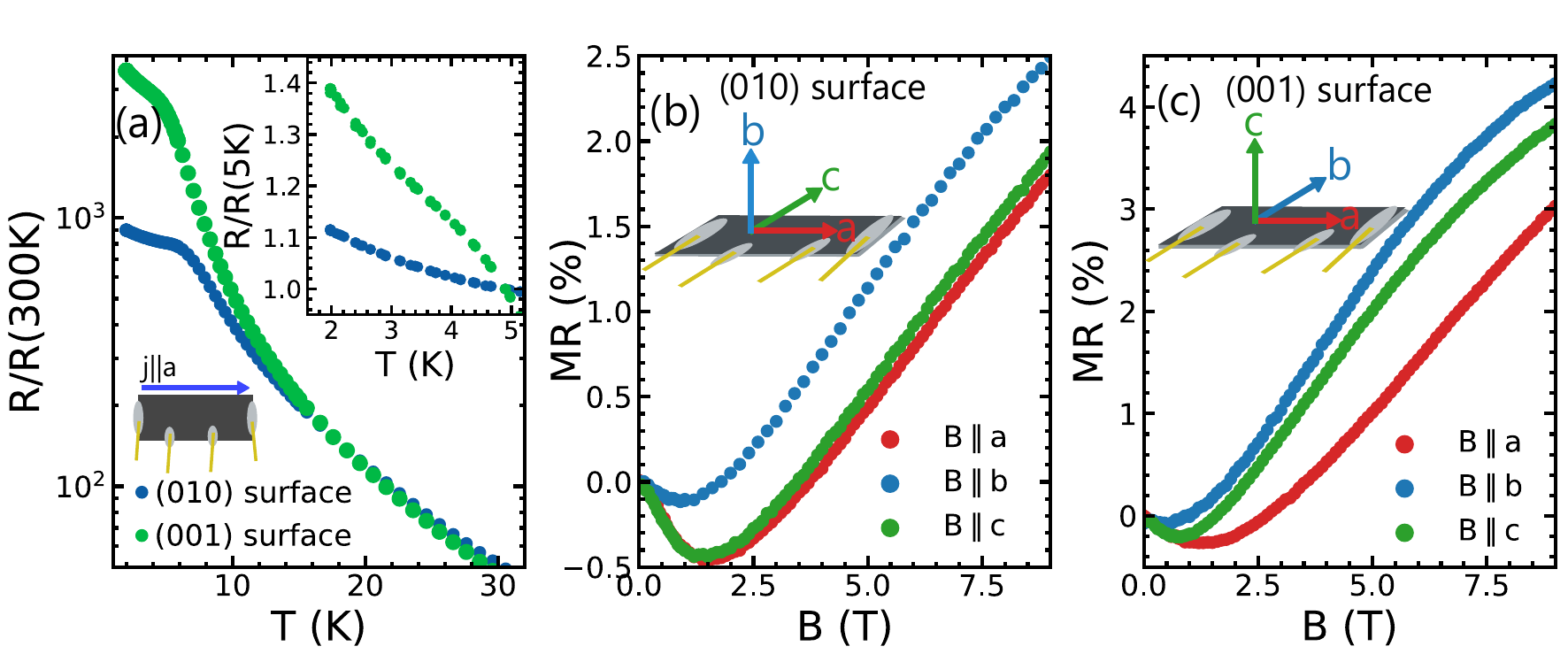}
    \caption{Comparison of the surface MR of uniaxial c-axis current geometry for transport primarily on [010] and [001] surfaces. (a) Temperature dependence of the 300K resistance ratio shows similar bulk transport character between the two samples with divergence below 10K with the transition to surface dominated transport. (b-c) Magnetic field dependence for field along a,b and c axes (or [100], [010] and [001], respectively) for current primarily on (b) (010) surface and (c) (001) surface.}
    \label{fig:010_vs_001}
\end{figure*}

For comparison with recent ARPES results, which show surface states on [010] surface and absence of surface states on [001] surface \cite{Li2024}, we examine the magnetotransport properties of these surfaces with uniaxial current direction along [100] direction.  Although contributions from both surfaces cannot be fully isolated in this transport geometry, a combined effort of preparing very thin and smooth surface with fine polishing was made in order to best compare the transport contributions from these surfaces for the same current and field directions. The [010] surface sample is 200 $\mu$m wide by 35 $\mu$m thin and the [001] surface sample is 400 $\mu$m wide by 20 $\mu$m thin. 
As shown in Fig.~\ref{fig:010_vs_001} (a),
the bulk-dominated transport (i.e., between 20-300 K) of these samples are nearly identical, confirming that the current direction is well aligned along the same direction for these two samples, since the bulk transport is well known to be anisotropic \cite{Petrovic2005}. 

However, as shown in the inset of Fig.~\ref{fig:010_vs_001} (a), the low-temperature surface-dominated transport is more insulating for the [001] surface sample than that of the [010] surface sample. We characterize the MR response of these surfaces by again applying field along each crystallographic axis and measuring the field response.
The measured anisotropy is qualitatively consistent  with the Corbino disk geometry samples, as shown in Fig.~\ref{fig:010_vs_001} (b) and (c) (i.e.,  the largest negative MR component is observed for  \textit{a}-axis field, while it is minimal or negligible for field along \textit{b}-axis).
Previous studies suggested that the observed low-field negative MR is a consequence of interference effects such as weak localization \cite{Takahashi2011}. In this scenario, magnetic flux through self-intersecting loops in coherent scattering paths gives rise to interference that make it more likely for electrons to travel in circularly localized paths \cite{Datta1995}. This effect, however, should strongly depend on the angle between the magnetic field and the motion of charge carriers. It is also far more prevalent in 2D systems due to the lower probability of self-intersecting loops in 3D. The results of Fig.~\ref{fig:010_vs_001} (b) and (c) are thus inconsistent with 2D weak localization, since self-intersecting loops perpendicular to the field along the current direction is forbidden in this transport geometry. As for three-dimensional weak localization, our results would imply an unlikely scenario in which interference effects are absent for scattering in the [010] plane but play a significant role for scattering in all other planes. 

\section{Low Temperature Hopping Conduction}

\begin{figure*}
    \centering
    \includegraphics[width=0.9\linewidth]{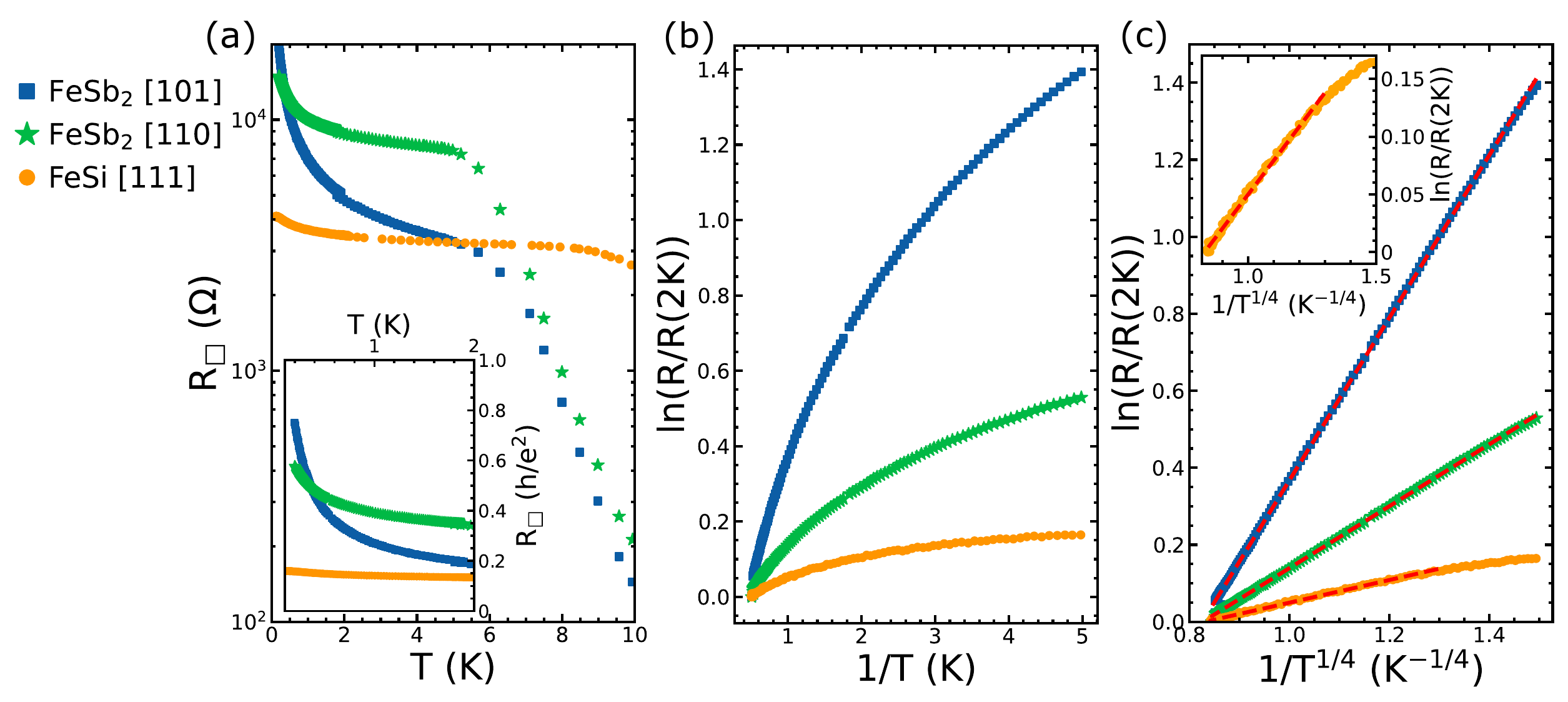}
    \caption{(a) Extended sheet resistance  measurement to low temperature on semi-log scale with inset linear scale in units of $h/e^2$ for FeSb$_2$ [101] (blue squares), FeSb$_2$ [110] (green stars) and FeSi [111] (orange circles). (b) Sublinear dependence of logarithmic resistance on inverse temperature. (c) Power dependence of logarithmic resistance on inverse temperature with exponent of $1/4$ fit (red dashed lines) to extract $T_{VRH}$ for each surface.}
    \label{fig:ADR}
\end{figure*}

The crossover from bulk to surface-dominated transport occurs around 10~K in FeSi and 5~K in FeSb$_2$, as shown by the plateau in $R(T)$ in Fig.~\ref{fig:ADR}a. On cooling through the surface-dominated regime, the resistances of both [110] and [101] surfaces, as well as for [111] FeSi, continue to increase. This is atypical of the expected behavior of metallic conduction. However, the increase is also inconsistent with simple thermal activation across a charge gap, as indicated by a sublinear (rather than linear) dependence of ln$(R)$ vs. 1/$T$ as shown in Fig.~\ref{fig:ADR}b.
Instead, the logarithmic resistance is a power law of temperature, consistent with variable-range-hopping (VRH) conduction as shown in Fig.~\ref{fig:ADR}c (although this power law in the FeSi sample deviates at lowest temperatures shown in the inset of Fig.~\ref{fig:ADR}c).  
 
Mott VRH typically describes transport in a heavily doped or amorphous semiconductor at low temperatures, where conduction is dominated by hopping between sites with a temperature-dependent length-scale. In this regime, charge carriers are localized and do not form extended states in the material (i.e. beyond the limit $k_Fl>1$). However, the inset of Fig.~\ref{fig:ADR}a shows that the sheet resistance from a geometric factor for a 2D surface Corbino disk (using dimensions in Fig.~\ref{fig:overview}a,b) does not exceed $h/e^2$,  the expected value for a 2D metal-to-insulator transition according to the Mott-Ioffe-Regal (MIR) limit \cite{DasSarma2014}. A possible explanation for this is that the assumption of a strictly 2D surface state, which was used to calculate the sheet resistance, is not valid and would therefore give a smaller effective sheet resistance due to the finite thickness of a 3D conduction channel.
 
The temperature dependence for VRH conduction is dimensionally dependent and takes on the form 

\begin{equation}
\label{eqn:Mott-VRH}
\rho (T) = \rho_0 \exp \left[ \left(\frac{T_{VRH}}{T} \right)^p \right],
\end{equation}

where $p = 1/(d+1)$ for a $d$-dimensional system (except $p=1/2$ for Coulomb-gapped systems) and $k_BT_{VRH}=\beta/(g(\mu)a^d)$ is the energy scale of the hopping channel with $g(\mu)$ density of states and hopping length scale, $\alpha$, and numerical coefficient, $\beta$ \cite{Shklovskii1984}. This exponential power law readily describes well-known systems, such as the cases of $p=1/4$ scaling of 3D VRH in p-type Si \cite{Shafarman1986}, $p=1/3$ scaling of 2D VRH in GaAs/Al$_x$Ga$_{1-x}$As quantum wells \cite{Qiu2012,Chakhmane2019} and the crossover from $p=1/d+1$ to $p=1/2$ in oxides with strong Coulomb interactions \cite{Rosenbaum1991} and heavily irradiated graphene \cite{Zion2015}.

A $p=1/4$ power law describes the logarithmic resistance well for the full temperature range below the crossover to the surface dominated regime, as shown in Fig.~\ref{fig:ADR}c, although the FeSi sample deviates towards more conducting behavior at lowest temperatures (for more details in finding the appropriate power law, see e.g. ref \cite{Zabrodskii2001}). This deviation is consistent with the observation that the surface of FeSi is the most conducting of the three samples when comparing sheet resistance in Fig.~\ref{fig:ADR}a. Also worth noting is that in FeSi samples grown by Sn-flux, the surface states show metallic conduction with no sign of hopping conduction \cite{Fang2018}, so the existence of parallel metallic conduction channels would not be unusual for this material. The temperature scale $T_{VRH}$ for hopping conduction in these three samples varies dramatically as shown in table \ref{table:VRH}, with $T_{VRH}$ for FeSi being over 2000 times smaller than that of FeSb$_2$ [101] surface, despite having a conductivity of the same order of magnitude.  

While it is surprising for an expectedly 2D surface state to exhibit 3D hopping character, the apparent contradiction is not unique. For example, Septianto {\it et al} reported a $p=1/4$ power law for VRH in a 2D superlattice of PbS quantum dots, a 2D nanoscale network exhibiting 3D-like conduction. Other apparent violations of the $p=1/3$ power law scaling in 2D systems have been described using an assumed temperature dependence of the resistance prefactor, $\rho_0$ \cite{Li2011,VanKeuls1997}. In the case of the disordered surface states of the topological Kondo insulator candidate, SmB$_6$, reported by Batkova {\it et al.}, a temperature independent parallel resistance channel was introduced to fit the data to a $p=1/3$ power law. As another example of a disordered surface state on a bulk 3D insulator, such results are important for understanding our current work. These deviations from $p=1/3$ scaling in 2D (or quasi-2D) systems are important for understanding the limits of the standard analysis for hopping conduction as well as how the effective transport length-scales compare with the physical dimensions (e.g. effective thickness) of these samples.

\begin{table}
\caption{Summary of low temperature hopping transport in FeSi and FeSb$_2$.}
{
\centering
\begin{tabular}{p{0.15\textwidth} p{0.1\textwidth} p{0.1\textwidth} p{0.1\textwidth}}
\hline\hline
Sample  &  T$_{VRH}$ (K) & R$_\square$ $(h/e^2)$ & R$_\square$ $(h/e^2)$ \\
  &   & T=2K & T=0.2K\\
\hline
 FeSb$_2$ $[101]$ & 19.8 & 0.348 & 0.751\\
 FeSb$_2$ $[110]$ & 0.416 & 0.197 & 0.578\\
 FeSi $[111]$  & 0.0075 & 0.135 & 0.158\\

\hline\hline
\end{tabular}\par
}
\label{table:VRH}
\end{table}

\section{Fitting MR to hopping and local moment scattering models}

\begin{figure*}
    \centering
    \includegraphics[width=0.95\linewidth]{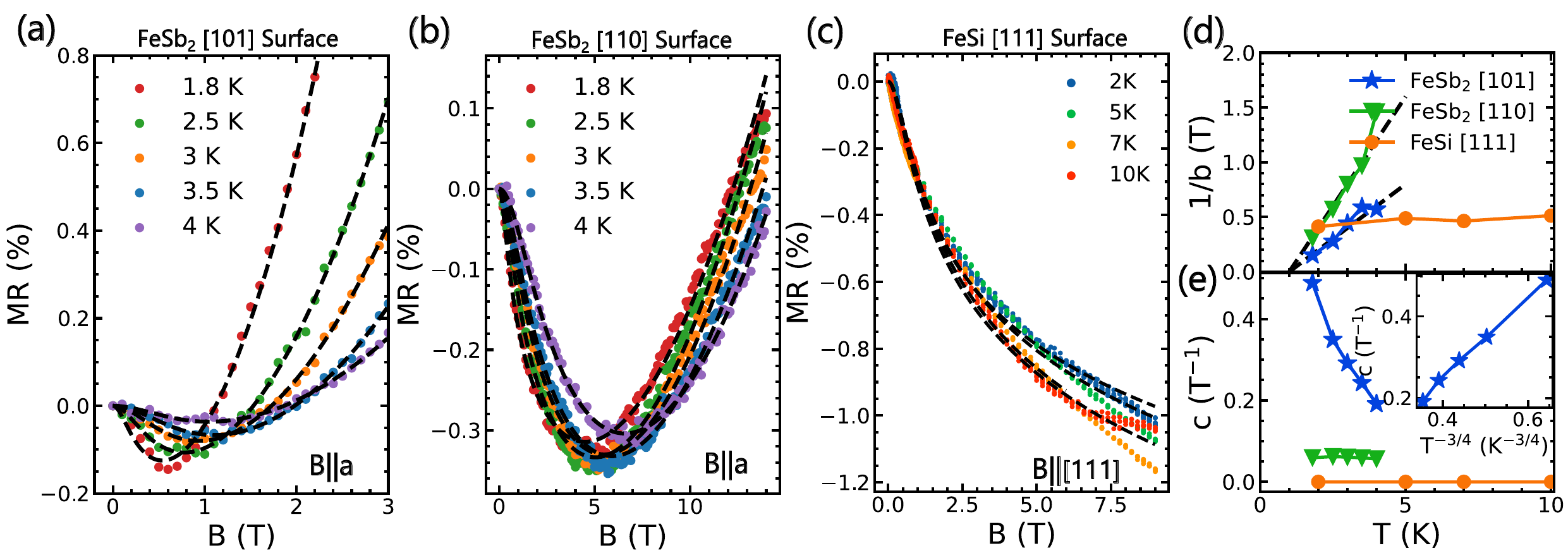}
    \caption{Results from fitting the MR of (a) FeSb$_2$ [101] surface MR with field along a-axis, (b) FeSb$_2$ [110] surface MR with field along a-axis and (c) FeSi [111] surface MR with field along [111] direction to equation \ref{eqn:Magnetoresistance} with black dashed lines representing lines of best fit. (d-e) Extracted values of the magnetic field scale of (d) local moment scattering and (e) suppression of hopping conduction. Inset of (e) shows FeSb$_2$ [101] fit results most closely resembles the expected temperature scaling for the field scale of hopping conduction suppression.}
    \label{fig:MR_fits}
\end{figure*}

The low temperature MR is inconsistent with conventional orbital (Lorentz force driven) MR as established by the angle dependence and negative MR at low fields. Similarly, the low temperature resistance is consistent with hopping conduction rather than metallic transport. To reconcile these observations, we first considered the effect of magnetic field on hopping conduction. In semiconductors doped towards their insulator-to-metal transitions like n-Ge and various III-V systems, a large, exponential dependence of resistance on magnetic field has been shown \cite{Sasaki1965, Sernelius1979, Monsterleet1997, Oiwa1998}. This effect is attributed to the shrinking of the spatial extent of localized states such that the overlap between neighboring impurity sites is reduced and therefore hopping conduction is suppressed \cite{Shklovskii1984}. This effect is largest for lightly doped semiconductors in which overlap at zero field is already small, so it is well understood to scale with the effective impurity radius and density of states at the Fermi level and takes the form

\begin{equation}
\label{eqn:Shklovskii_Magnetoresistance}
\left( \frac{\rho(H)}{\rho(0)} \right)_{VRH} = \exp\left[ \frac{tae^2}{Nc^2\hbar^2}\left(\frac{T_{VRH}}{T}\right)^{3/4}H^2 \right]
\end{equation}

where $a$ is the impurity site radius, $N$ is the impurity concentration and $t=5/2016$ is a numerical coefficient \cite{Shklovskii1984}. This field dependence only accounts for the increase in resistance, while in some hopping conduction systems (and in the data we present), a decrease in resistance at low fields is observed. In some heavily doped semiconductors, negative MR has been attributed to the Zeeman effect in which carriers in the upper Zeeman state have an increased wavefunction overlap with neighboring states. This effect was proposed by Filiuama and Yosida (see Ref.~\cite{Fukuyama1979}) and used to describe the low field MR of n-type GaAs \cite{Benzaquen1988}. However, this effect should scale with temperature proportional to $T^{1/4}$ (i.e. decreasing with cooling) in contradiction with FeSb$_2$ results showing a negative MR component increasing with cooling as shown in Fig.~\ref{fig:MR_fits}. Negative MR in variable range hopping systems are typically attributed to quantum interference from self-intersecting loops within paths connecting hopping sites \cite{Shklovskii1984,Abdia2009}. This interference, as discussed in a previous section, should depend on field direction relative to current, which is absent in the MR of FeSb$_2$. 

A possible explanation for the anisotropic negative MR observed in FeSb$_2$ is the effect of local moment scattering, which could be anisotropic with respect to the crystal field. The usual $s$-$d$ scattering picture of conduction electrons coupling to local moments works well for understanding the origin of negative MR in metals with magnetic impurities. Although the validity of applying $s$-$d$ scattering models to VRH systems is not straightforward, experimental evidence of the coupling of transport to local magnetic moments in VRH systems is clearly observed for dilute magnetic semiconductors, and even at the surface of FeSi \cite{VanEsch1997,Park2001,Avers2024}. Therefore, we consider the contribution of magnetic impurities to the negative MR of our samples at low temperatures. 

 Bulk FeSb$_2$ does not exhibit large magnetocrystalline anisotropy, as it's ground state is singlet \cite{Petrovic2005,Li2024}. However, when electron-doped by Te substitution into a bulk metal, for example, the magnetic properties become extremely anisotropic with [100] magnetic easy axis \cite{Hu2009}. Local moment scattering in metals, from second order perturbation theory, should give a $-M^2$ (for $M$ magnetization of local moment) contribution to MR at small magnetic fields and thus follow a Curie-Weiss scaling with temperature for negative quadratic MR. Although this scaling has been shown to work well to describe Cu-Mn alloys \cite{Yosida1957,Toyozawa1962}, it is limited and incomplete in describing local moment scattering for some systems in which impurity scattering from local moments plays a role. A calculation based on the third-order expansion of the $s$-$d$ exchange interaction is used here, adapted by Khosla and Fischer \cite{Khosla1970} from early theory work on tunnel junctions with transition metal impurities \cite{appelbaum1967}:

\begin{equation}
\label{eqn:Khosla-Fischer}
\left( \frac{\Delta\rho(H)}{\rho(0)} \right)_{KF} = -A_1J\rho_F \ln \left( 1+ \left(\frac{g_0\mu}{\alpha k_BT}\right)^2B^2 \right)
\end{equation}

The original Khosla and Fischer paper also gives a quadratic positive MR contribution in addition to this negative component from local moment scattering. The authors make it clear that a Lorentz force contribution is not responsible for the observed positive MR as they observe positive longitudinal MR of the same order as transverse, similar to the data we present in this paper. They suggest a possible two-band scenario in which the relative band populations, and therefore the effective resistance, change with field to give an overall positive quadratic component. However, in the heavily doped regime (in which $N/a$ is large) or at small fields, Eq.~\ref{eqn:Shklovskii_Magnetoresistance} can be expanded to first order in small argument to give a quadratic MR which is consistent with the observed MR of Khosla and Fischer. Giving a complete expression for carriers hopping in a doped semiconductor, in which local moments play a leading role in the scattering of charge carriers, we have:

\begin{equation}
\label{eqn:Magnetoresistance}
\frac{\rho(H)}{\rho(0)} = -a\ln(1+b^2B^2) + \exp(c^2B^2),
\end{equation}

where the field scales of local moment scattering (i.e., $b$ from Eq.~\ref{eqn:Khosla-Fischer}) should scale with $T^{-1}$, and field scales of local wavefunction shrinkage (i.e., $c$ from Eq.~\ref{eqn:Shklovskii_Magnetoresistance}) should scale with $T^{-3/4}$.

The temperature dependence of MR in the surface transport regime agrees well with the expected behavior from Eq.~\ref{eqn:Magnetoresistance} for the [101] surface of FeSb$_2$ as shown in Fig.~\ref{fig:MR_fits}. The field scale of the negative MR component evolves linearly with temperature while the field dependence of the hopping contribution (positive MR component) scales as $T^{3/4}$. The temperature evolution of the hopping field scale, $c$, for the FeSb$_2$ [110] and FeSi [111] surfaces appear to be inconsistent with the expected scaling in this temperature range ($T\geq2 K$). This can be attributed to the small hopping temperature relative to the temperature range being measured in Fig.~\ref{fig:MR_fits}, such that the surface carriers are too delocalized to be sensitive to the small changes in localized state wave-functions with applied field. For both FeSb$_2$ samples, we find good agreement with the model from Khosla and Fischer, which does not have explicit dependence on the nature of conduction. The FeSi sample, however, shows no agreement with the temperature dependence expected from the Kholsla and Fischer model. This isn't surprising since there is a built-in assumption of non-interacting local moments in the model, which contrasts with the reported magnetic order at the surface of FeSi that plays a significant role in the observed negative MR \cite{Avers2024,Ohtsuka2021,Deng2023}.

\section{Discussion}
The transport character of our FeSb$_2$ and FeSi samples is consistent with a disordered surface conduction layer with an effective thickness larger than the characteristic length-scale of the hopping channel. This result provides good evidence that FeSb$_2$ and FeSi are not strong topological insulators, since our transport results show conduction by 3D localized states unlike the 2D metallic (extended) states expected at the surface of a strong topological insulator. Without knowing the dielectric properties of FeSb$_2$, it is difficult to estimate the length scale of the hopping channel, $l_{hop}$, but it may be as small as a few or tens of nanometers and decreases with temperature as $l_{hop} \propto a_{eff}(\frac{T}{T_{VRH}})^{1/4}$ for some effective Bohr radius, $a_{eff}$~\cite{Shklovskii1984}. While the value of $T_{VRH}$ varies by orders of magnitude among these samples, it should be noted that $l_{hop}$ may only vary by a factor of 2.6 in between the [110] and [101] surfaces of FeSb$_2$ and a factor of 7.2 between FeSb$_2$ and FeSi, assuming the same value of $a_{eff}$ (the value of which depends on band structure and, in mixed-valence insulators, can be orders of magnitude smaller than expected based on effective mass alone \cite{Skinner2019}). 

The origin of this surface transport channel is not fully understood. Based on the crystalline anisotropy found in the surface MR, and the fact that spectroscopic observations of surface states were measured on surfaces cleaved in vacuum, a disordered conducting oxide layer can be ruled out. A recent ARPES and ab-initio combined study by Chikina {\it et al.} concluded that small structural distortions of a few percent of the bulk lattice parameters is enough to shift a bulk band near the Fermi level to match observed surface state dispersion results \cite{Chikina2020}. While only speculation, this would suggest that these surface states appear due to some weak structural distortion that is not present in the bulk. 

In a series of experiments investigating structural defects via scanning transmission electron microscopy (STEM) and neutron scattering measurements, Du {\it et al.} reported structural distortions due to both Fe and Sb vacancies in all measured samples of FeSb$_2$ with various growth conditions \cite{Du2021}. These vacancies are frequent enough to show site occupations as small as 82\% in some Sb sites and 94\% in some Fe sites. It was reported that samples with these levels of vacancies also show a monoclinic distortion that is evidenced by ``forbidden reflections" in neutron scattering, indicating some reduced symmetry. It is worth noting that, while Du {\it et al.} suggest these defects contribute in-gap impurity states that modify bulk physical property measurements, recent measurements of the bulk resistivity using the inverted resistance method show no sign of bulk impurity conduction \cite{Eo2023}. This implies that defects present in the bulk may  contribute to scattering processes but do not form an impurity band, and that somehow the bulk electronic gap is incredibly robust to these defects. Therefore, the presence of trivial surface states seems to suggest that either the robustness of the bulk breaks down or the structural distortions due to vacancies is stronger near the surface.

\section{Conclusion}
Magnetotransport measurements of two isolated surfaces of the candidate $d$-electron topological Kondo insulator FeSb$_2$ were measured using both Corbino and oriented four-wire geometries. Our measurements of field-angle-dependent low-temperature electrical transport properties point to the following conclusions:  

\begin{enumerate}[label=(\roman*)]
\item a crossover from orbital-dominated bulk MR at high temperatures, to a surface state angular response dominated by crystalline anisotropy at low temperatures;

\item a low temperature negative MR that is inconsistent with weak localization;

\item electrical transport properties sensitive to the conducting surface, with (010) and (001) comparisons consistent with ARPES results \cite{Li2024};

\item a surface hopping transport behavior best described by power laws consistent with 3D conduction;

\item a combined hopping and local moment scattering model best describes the temperature and magnetic field dependence of the surface MR.

\end{enumerate}

We show that these results are qualitatively similar to the surface transport properties measured on the [111] surface of FeSi, which also exhibits low temperature three-dimensional variable range hopping conduction and magnetic scattering. This suggests that the surface states of FeSb$_2$ are best described as a disordered semiconducting layer of finite thickness, instead of a two-dimensional electron gas as expected for the surface states of a strong topological insulator.

\bibliography{references.bib} 
\end{document}